\documentclass[journal,twoside,web]{ieeecolor}
\usepackage{tmi}
\usepackage{hyperref}
\usepackage[utf8]{inputenc}
\usepackage[T1]{fontenc}
\usepackage{color,soul}
\usepackage{times}
\usepackage{amssymb} 
\usepackage[nointegrals]{wasysym} 
\usepackage{multicol,multirow}
\usepackage{booktabs}
\usepackage{mathtools}
\usepackage{subcaption}
\usepackage{cite}
\usepackage{amsmath,amssymb,amsfonts}
\usepackage{algorithmic}
\usepackage{graphicx}
\usepackage{textcomp}
\usepackage{adjustbox}

\begin{document}
\title{Advancing Medical Image Segmentation with Mini-Net: A Lightweight Solution Tailored for Efficient Segmentation of Medical Images}


\author{Syed~Javed, Tariq~M.~Khan, Abdul~Qayyum, Hamid Alinejad-Rokny, Arcot~Sowmya,Imran~Razzak
\thanks{S.Javed, T.M.Khan, A. Sowyma, I. Razzak are with the School of Computer Science and Engineering, University of New South Wales, Sydney, NSW, Australia (e-mail: \{s.jave, tariq.khan, a.sowmay, imran.razzak\}@unsw.edu.au).}
\thanks{A. Qayyum is with National Heart and Lung Institute, Faculty of Medicine, Imperial College London, London, United Kingdom (e-mail: a.qayyum@imperial.ac.uk).}}%


\maketitle              
\begin{abstract}

Accurate segmentation of anatomical structures and abnormalities in medical images is crucial for computer-aided diagnosis and analysis. While deep learning techniques excel at this task, their computational demands pose challenges. Additionally, some cutting-edge segmentation methods, though effective for general object segmentation, may not be optimised for medical images. We propose Mini-Net, a lightweight segmentation network specifically designed for medical images to address these issues. With fewer than 38,000 parameters, Mini-Net efficiently captures both high- and low-frequency features, enabling real-time applications in various medical imaging scenarios. We evaluate Mini-Net on various datasets, including DRIVE, STARE, ISIC-2016, ISIC-2018, and MoNuSeg, demonstrating its robustness and good performance compared to state-of-the-art methods.

\end{abstract}

\section{Introduction}

Medical image segmentation represents a cutting-edge convergence of medical imaging and computer vision, with a focus on extracting meaningful insights from intricate medical images. The surge in imaging technologies such as magnetic resonance imaging (MRI), computed tomography (CT), and PET underscores the growing importance of accurately delineating and analysing anatomical structures or pathological regions within these images. This precision has become indispensable in clinical diagnosis, treatment planning, and medical research.

Accurate segmentation of anatomical structures and abnormalities in medical images is essential for a precise diagnosis and optimal treatment planning \cite{imtiaz2021screening,soomro2018impact,trinh2024sight,tang2024discriminating}. However, this task poses significant challenges, even for human experts, due to factors such as ambiguous structural boundaries, diverse textures, imbalanced intensity distribution, inherent uncertainty in segmented regions, contrast variations, and scarcity of annotated datasets. The urgency of automated segmentation techniques in medical imaging has spurred numerous research endeavours aimed at overcoming these challenges. For example, a fully convolutional multiscale residual network was proposed for segmentation of retinal vessels, using three multi-scale kernels to capture large, medium, and thin vessels\cite{khan2021residual}. Segmentation of large and thin retinal vessels was addressed through a block matching mechanism and multiscale triple stick filtering\cite{khan2022width}. An improved ensemble block matching was also proposed to automate the detection of fine vessels in noisy fundus images\cite{naveed2021towards,de2023airogs}. Existing segmentation techniques can be broadly categorised as supervised and unsupervised approaches. Supervised approaches involve learning from annotated training images provided in pairs (image, mask), whereas unsupervised methods lack annotation and rely on low-level features and ad-hoc rules, which limit their generalisability.

Supervised deep learning-based techniques, particularly convolutional neural networks (CNN), have emerged as leaders in medical image segmentation\cite{mazher2024self,qayyum2023two,qayyum2024unsupervised,qayyum2023semi}. Despite the prowess of these models, there is a need for solutions tailored to resource-constrained devices. To meet this challenge, Khan et al. \cite{khan2022leveraging} analysed image complexity to develop a macrolevel neural network for medical image segmentation. They use a variant of U-Net with a decreased number of filters and reduced depth of encoder blocks to minimise the model capacity and size. Iqbal et al. \cite{iqbal2022g} devised a small-scale neural network for the segmentation of retinal vessels, eliminating feature overlap to reduce computational redundancy.  \cite{khan2022mkis} refines the receptive field using multiple kernels with different sizes to improve segmentation performance. \cite{khan2023simple} utilises a multi-scale cascaded path to design a network with 1.3 million parameters for polyp segmentation. 
In \cite{khan2023feature}, the authors present a feature enhancement segmentation network that alleviates the need for pre-training image enhancement, reducing associated computational overhead. The authors of \cite{iqbal2023ldmres,khan2023esdmr} and \cite{mobilenetv3} build networks with a restricted number of trainable parameters, tailored for devices with limited resources. Although MobileNet-V3 \cite{mobilenetv3} excels in object segmentation, it is not optimised for medical image segmentation. In this paper, we introduce a remarkably lightweight model, Mini-Net, explicitly designed for medical image segmentation that caters to devices with limited computing power. Key contributions of this work include the following:
\begin{itemize}
    \item An innovative simplified architectural design consisting of dual multi-residual block (DMRes) and Expand Sequaze blocks tailored for medical image segmentation, incorporating robust features selection.
    \item The lightweight segmentation network (Mini-Net) is aided by a dual multi-residual block consisting of only $38k$ parameters, which beats all existing works and is super fast and memory efficient compared to existing models.
    \item  Extensive experiment conducted on multiple medical imaging datasets showed significant performance of the model, demonstrating state-of-the-art results.
\end{itemize}

\section{Literature Review}
Medical image segmentation has attracted the attention of researchers due to increased health complications and increased diseases due to environmental changes and lifestyles of people. Accurate segmentation of medical images poses significant challenges due to factors such as ambiguous structural boundaries, diverse textures, imbalanced intensity distribution, inherent uncertainty in segmented regions, contrast variations, and scarcity of annotated datasets. We will further discuss how researchers have attempted to meet these challenges. Existing segmentation techniques can be broadly categorised as supervised and semi-supervised approaches. In this section, we will discuss various aspects of medical image segmentation applications devised by different deep learning and computer vision specialists over the years.

\subsection{Supervised Deep Learning based Techniques}
Supervised deep learning-based techniques achieved the best results so far for the segmentation of images including medical images. There has been a notable improvement in neural network architectures for medical image segmentations in terms of model backbones, model building blocks, hyperparameters, and optimised loss functions. In semantic segmentation of medical images, we aim to classify every individual pixel in the image, and to achieve this, most researchers have proposed the encoder-decoder architecture that has been used in most of the current state-of-the-art techniques for segmentation such as U-Net \cite{ronneberger2015u}, generative adversarial networks (GANs) and numerous variants of U-Net. In encoder-decoder-based techniques, we have an encoder that extracts image features at various levels, and then the decoder blocks decipher the extracted features and restore the original image. 
The journey of supervised learning-based segmentation begins with fully convolutional neural networks (FCN). FCN was initially introduced by adding fully connected layers at the end of convolutional neural networks to obtain probability information. This was only for image classification and not for pixel-level classification. SegNet \cite{Badrinarayanan2017}, introduced by Nakazawa et al., is designed for pixel-level classification of images (i.e. segmentation) and is built upon the FCN semantic segmentation task and has an encoder-decoder-based structure. The authors use VGG16 as the network encoder block to retrieve image features, and the decoder block uses these features to assign a colour label to each pixel in the image. While FCN upsamples the low-resolution features with deconvolution operations, SegNet upsamples them using a more extensive pooling index from the encoder instead of learning how to do so. In this way, SegNet creates dense features using trainable convolution kernels on sparse feature maps, and the softmax classifier categorises pixels after restoring the maps to their original resolution. Unpooling of the low-level features maintains high-frequency data, which helps to preserve image details. This process can contribute to better performance in tasks that require fine-grained information, such as edge detection. Despite the advantages that SegNet offers, it also comes with challenges and limitations such as requiring resources with large memory and high computational power, overfitting, shallow semantic understanding, unable to handle occlusions and object interactions, producing noisy and jagged boundaries for objects, and having limited generalisation capability. We will need to take further precautionary steps to overcome the limitations of SegNet.

SegAN \cite{xue2018segan}, the adversarial segmentation network, is a U-Net-based network that uses adversarial learning for segmentation. The authors efficiently tackle the issue of class imbalance between pixel categories by alternatingly training a segmenter and a critic network in a Min-Max game and by using a multiscale L1 loss function. The multiscale L1 loss function helps capture both local and global features during training and consequently improves the segmentation performance of the network. Where adversarial learning and the multiscale L1 loss function improve the segmentation performance, they also come with enhanced complexity, making the network require more memory and computation power. This hampers the scalability of the model and its practical applicability in real time. The authors evaluate and discuss SegAN performance in BRATS2013 and BRATS2015 and do not discuss its applicability to any other medical datasets, nor do they say if the proposed methodology is generalisable in different medical applications.
The three-stage FCN \cite{8476171}, proposed by Yan et al., focusses on accurately segmenting retinal vessels in medical images. It employs a multistage architecture to progressively refine segmentation results, with the aim of improving accuracy and reducing false positives and false negatives. Like other deep learning-based techniques previously discussed, the three-stage FCN is computationally complex and costly. This model requires a large dataset for training, which is not available in the case of medical images. 

The "BTS-DSN" model proposed by Guo et al. \cite{GUO2019105} aims to perform retinal vessel segmentation using a deep-supervised neural network with short connections. The model employs a deeply supervised learning approach, which involves adding auxiliary supervision signals at intermediate layers of the network, which helps facilitate gradient flow during training and can lead to more stable convergence and improved segmentation performance. Furthermore, BTS-DSN uses short connections within the neural network architecture, which can help propagate information across different layers more effectively, aiding in feature extraction and segmentation accuracy. The authors use DRIVE, CHASEDB1 and STARE datasets to evaluate the proposed method and use data augmentation to enlarge the datasets. They have used traditional augmentation techniques, including rotation, flipping, and scaling, but do not mention the scaling size and reason. They train the network with a learning rate of $1e^{-8}$ that is rarely practised with a very minor learning rate decay. They do not mention why they chose these hyperparameters. Although the most commonly used learning rate that has resulted very well is 1$e^{-4}$. The authors also use ResNet-101 as the backbone, which causes the model to have a large capacity and to be computationally complex and costly. 

U-Net revolutionizes conventional CNN networks' application in medical image segmentation by adopting symmetrical structure skip connections and displaying state-of-the-art performance in image segmentation tasks. This strategic design overcomes specific challenges posed by medical images, including noise and unclear boundaries, while efficiently integrating low-level and high-level image features essential for precise segmentation in medical tasks. As a result, the U-Net stands out as the premier choice for medical image segmentation, catalyzing numerous breakthroughs in the field. Given the volumetric medical data like CT and MRI images that are in 3D format, researchers have ventured into extending U-Net's capabilities to 3D data. Çiçek et al. \cite{cciccek20163d} started with the 3D U-Net, specifically tailored for handling 3D medical data. However, the 3D U-Net's restricted depth, owing to computational limitations, compromises its capacity to capture intricate features, thus constraining segmentation accuracy. In response to this challenge, Milletari et al. \cite{milletari2016v} introduced the V-Net, a variant architecture integrating residual connections for deeper network structures. This innovation not only addresses issues like the vanishing gradient but also facilitates deeper architectures, thereby enhancing feature representation and segmentation performance. 
After the transformer’s enormous success on language models and its remarkable performance in vision applications, researchers were interested in merging the power of U-Net with transformer and many transformer-based U-Net models such as Trans-UNet \cite{chen2021transunet}, Swin-UNet \cite{cao2023swin} and UNet++ with Vision Transformer were proposed. Whereas standard U-Net fails to capture global features effectively, transformer-based U-Net models address this limitation by replacing the convolutional layers with transformer blocks in the standard U-Net encoder. This self-attention mechanism helps the model to capture long-range dependencies efficiently, leading to overall improved segmentation performance.
Proposed by Zhou et al. U-Net++ \cite{zhou2018unet++} aims to address some limitations of the standard U-Net model in capturing multi-scale contextual information efficiently. U-Net++ presents notable strengths in image segmentation with its ability to enhance accuracy through nested skip connections, capturing multi-scale contextual information, and deep supervision mechanisms, which facilitate learning features at various abstraction levels. This hierarchical feature learning capability enables the model to effectively segment complex structures in images. However, these advantages come with limitations. The increased computational complexity of U-Net++, stemming from its deeper architecture and dense connectivity, can pose challenges during both training and inference, potentially demanding substantial computational resources. Additionally, training U-Net++ requires more time and careful optimization due to its complexity, and there is a heightened risk of overfitting, especially with limited training data. Interpretability may also be compromised by the dense connectivity, and the model may require more memory resources during deployment, which could be problematic in resource-constrained environments like edge devices or real-time applications.
The improved performance of the different variations of U-Net is undeniable, yet they come with the challenges of increased computational complexity, excessive memory requirements, and high chances of overfitting as compared to standard U-Net. Besides these challenges, transformer-based U-Net models require vigilant optimization and tuning of hyperparameters because of their hybrid nature and large parameter space.

\subsection{Semi-supervised Deep Learning based Techniques}
In this area of research, the goal is to efficiently address the challenge of limited annotated data by using both labelled and unlabelled medical images for the training of segmentation models. This approach specifically suits medical images as there is always a shortage of annotated dataset that is large enough for the application. Semi-supervised segmentation is a common scenario in medical applications where a small portion of the training images are annotated, while we also have a large unannotated portion that can be used to improve both the accuracy and generalisation capability of the model. Several algorithms and models have been proposed in this area to reduce the cost of labor-intensive, pixel level annotations of large medical images datasets. 

One of the common ways to deal with limited annotated dataset is data augmentation and the most used augmentation technique is the traditional parametric transformation of images such as translation, scaling, shifting, rotation, horizontal and vertical flips, etc. In addition to the traditional augmentation technique, researchers have also used conditional generative adversarial networks (cGANs) for the augmentation and synthesis of medical images. Several works, including \cite{shin2018medical, jin2018ct} have used these augmentation techniques to enlarge the dataset and improve the model performance. The authors in \cite{shin2018medical}, introduce a way to synthesise medical images using GANs that can help anonymise sensitive medical data. However, the quality of the synthesised images is questionable, since GANs can struggle to generate images with the level of detail and fidelity required for medical applications. The paper does not provide sufficient evaluation and validation of the method on clinical datasets, making it difficult to assess the performance of the proposed method in capturing accurate anatomy and pathology. Although \cite{jin2018ct} produces synthetic data that closely resembles real-world CT scans, facilitating more realistic and clinically relevant evaluations of lung segmentation algorithms, they also fail to adequately address the realism and fidelity of synthetic nodules compared to real-world CT scans. Because cGANs generate images with blurred boundaries and low resolution, researchers have used CycleGAN to improve the quality of the synthesised images.
Another efficient way to deal with limited annotated data using semi-supervised learning is the transfer learning mechanism. In this setting, the trained and learnt weights of a pre-trained network are used to fine-tune a network on a new set of data with limited number of annotated and labelled samples. Researchers discovered that using pre-trained networks on natural images as an encoder for the U-Net like model and fine-tuning it on medical images improves the performance of the model for segmentation as well as classification tasks.

\subsection{Lightweight Medical Image Segmentation Models}
Following the success of lightweight models like MobileNet \cite{mobilenetv3} in general object segmentation, there has been growing interest among researchers in designing efficient, lightweight networks for medical image segmentation. The main focus has been to minimize network size and capacity, reduce the computational burden, and lower memory requirements. Iqbal et al. \cite{iqbal2023ldmres} introduced LDMRes-Net, a compact and efficient model built using dual multiscale residual blocks, which integrate a multiscale feature extraction mechanism. This allows the network to capture details at various granular levels, while also reducing the number of parameters and computational complexity compared to traditional deep learning models. The use of depth-wise separable convolutions further enhances the efficiency of LDMRes-Net, with residual connections ensuring that performance remains strong. Similarly, Khan et al. \cite{khan2023esdmr} proposed a lightweight network tailored for medical image segmentation, focusing on the capture of high-frequency features crucial for such tasks. Their model incorporates expand-and-squeeze blocks, which increase computational efficiency and robustness, making it suitable for deployment on devices with limited processing power. Li et al. \cite{lightweightUnet} introduced a lightweight version of U-Net for lesion segmentation in ultrasound images. This model balances computational efficiency and accuracy, making it a strong choice for applications where resources are constrained. An additional example comes from Ma et al. \cite{ma2018shufflenet}, who proposed ShuffleNet V2, a lightweight network known for its superior performance in mobile and embedded device scenarios. By employing a channel split operation, ShuffleNet V2 achieves an optimal balance between speed and accuracy, making it well-suited for tasks involving limited computational power. 

Despite these advances, there has been limited work on the development of lightweight models for medical image segmentation that works fine with general medical images. In this paper, we aim to address this gap by proposing a lightweight model for segmentation of medical images including retinal vessels, skin lesion and multi-organ nuclei, while maintaining state-of-the-art performance. This model will be optimized to work effectively on devices with limited computational resources, making it a valuable contribution to the field of medical image analysis.

\begin{figure*}[!htb]
\includegraphics[width=\textwidth]{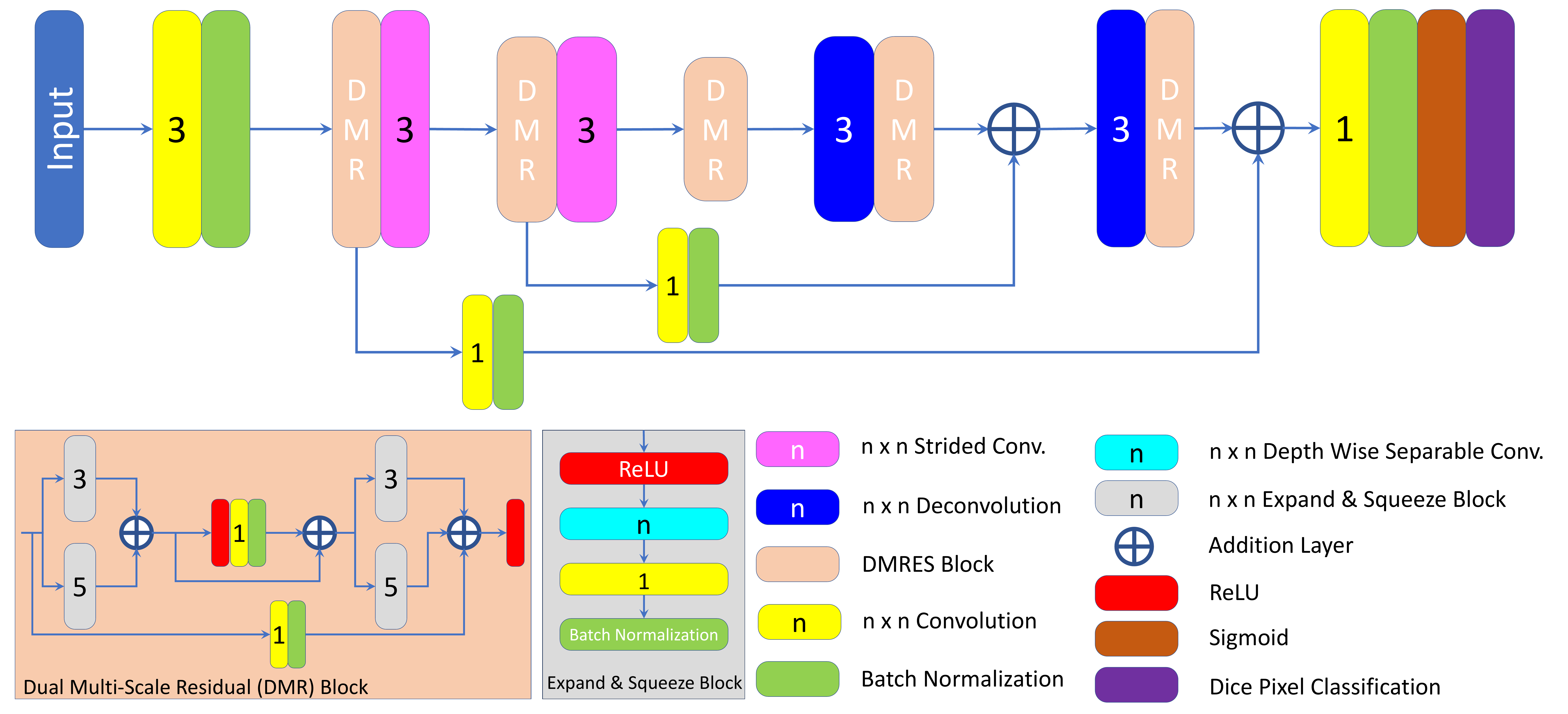}
\caption{Mini-Net Model Diagram} \label{model_diagram}
\end{figure*}

\section{Methodology}
We introduce Mini-Net, which is designed as a lightweight encoder-decoder model specifically crafted for the segmentation of medical images. Central to its architecture is the integration of a dual multiresidual block (DMRes) and an Expand Squeeze block, inspired by recent advances in feature extraction and regularisation techniques \cite{iqbal2023ldmres} and \cite{khan2023esdmr}. Mini-Net aims to strike a balance between capturing high-level semantic features and preserving fine-grained details inherent in medical imaging data. This balance is crucial for accurate segmentation, particularly in tasks involving anatomical structures or pathological regions.

\subsection{Mini-Net Architecture}
The architecture of Mini-Net is characterised by an encoder-decoder framework, with the DMRes block serving as its central component. Unlike traditional encoder-decoder models, Mini-Net places special emphasis on efficient feature extraction, achieved through the integration of DMRes blocks within the encoder pathway. These blocks facilitate multiscale feature extraction and refinement, enabling the model to capture both global context and local details present in the input images. This feature is particularly beneficial in medical imaging, where precise delineation of structures is paramount.

 Figure \ref{model_diagram} shows the diagram of the Mini-Net model. The input of the model denoted $X_{in}$, is represented as a three-dimensional tensor with dimensions $C \times H \times W$, where $C$ represents the number of channels and $H$ and $W$ denote the height and width of the input image, respectively. The operation $f^{n \times n}(\cdot)$ denotes a convolution operation with a kernel size of $n \times n$, and $\beta_{n}(\cdot)$ represents batch normalisation.

The initial feature map, denoted $f_{1}$, is obtained by processing the input image $X_{in}$ through a convolution operation followed by batch normalisation, as expressed in Equation \ref{Eq1}:

\begin{equation}
    f_{1}=\beta_{n}\Big(f^{3\times 3}(X_{in})\Big)
    \label{Eq1}
\end{equation}

The feature map, $f_{1}\in\mathbb{R}^{\hat{C}\times H \times W}$, is then fed as input to the first encoder block. Each encoder block has a DMRes block followed by a strided convolution operation. So, $f_{1}$ is fed into the DMRes block where multi-scale feature extraction and feature refinement are performed. The $S_{out_1}^{dmr}$ is the output of the DMRes block given in (Eq. \ref{Eq:mult3}), where $k=2\times i+1$.

\begin{equation}
    S_{1}^{dmr}=\Re\left ( \sum_{i=1}^{2}\beta _{n}\Big ( f^{k\times k} ( f_{1} ) \Big ) \right ) 
    \label{Eq:mult1}
\end{equation}

\begin{equation}
    S_{2}^{dmr}=\Re\Bigg(  \beta_{n}\Big(f^{1 \times 1}(S_{1}^{dmr})\Big) + S_{1}^{dmr} \Bigg)
    \label{Eq:mult2}
\end{equation}

\begin{equation}
    S_{out_1}^{dmr}=\beta _{n}\Big(f^{1 \times 1}(f_{1})\Big)+\sum_{i=1}^{2}\beta _{n}\Big(f^{k\times k}(S_{2}^{dmr})\Big)
    \label{Eq:mult3}
\end{equation}

$S_{2}^{dmr}$ and $S_{1}^{dmr}$ are the intermediate outputs of the addition layers of the dual multiscale residual block and are calculated as (Eqs. \ref{Eq:mult2}-\ref{Eq:mult1}). We have used convolution operations with kernel sizes $1 \times 1, 3 \times 3$ and $5\times5$ to obtain features on multiple scales and then added residual connections to maintain high-frequency features. Now that we have feature maps, $S_{out_1}^{dmr}$, achieved from the DMRes block, we feed it into the strided convolutional layer of the encoder block, $f_s^{n\times n}$, where $n$ is the kernel size, for downsampling of the feature maps as computed in (Eq.\ref{str1}). 

\begin{equation}
    \Im _{1}^{enc}=f_s^{3\times 3}(S_{out_1}^{dmr})
    \label{str1}
\end{equation}

Here $\Im _{1}^{enc} \in\mathbb{R}^{(\hat{C}\times 2)\times \frac{H}{2}\times \frac{W}{2}}$, where we take $\hat{C}$ as 8, is the output of the first encoder block which is fed as the input to the second encoder block where the same sequence of steps is followed as outlined in (Eq. \ref{Eq:mult1}-\ref{Eq:mult3}). The output of the second DMRes block, $S_{out_2}^{dmr}$, is further fed into the second decoder that generates $\Im_2^{enc}\in\mathbb{R}^{(\hat{C}\times 4)\times \frac{H}{4} \times\frac{W}{4}}$. This value is then directed to the bottleneck block, which comprises a single DMRes block that yields the final output of the encoder blocks, $\Im_{out}^{enc}\in\mathbb{R}^{(\hat{C}\times 4)\times\frac{H}{4}\times\frac{W}{4}}$. This output is now ready to be fed into the first decoder block. It is essential to note that in the bottleneck, we solely refine the feature maps while maintaining the same spatial dimensions as $\Im_ 2^{enc}$.

Our decoder blocks mirror the architecture of the encoder blocks, initiating with deconvolution operations for up-sampling, succeeded by DMRes blocks. The initial decoder begins with a deconvolution layer, as delineated in (Eq. \ref{Eq:dec1}). Subsequently, the output of the decoder blocks is calculated according to the formulations in (Eq.\ref{Eq:mult6}-\ref{Eq:mult4}), where $f_d^{n\times n}$ denotes a deconvolution operation with a kernel size of $n\times n$.

\begin{equation}
     D_{1}^{dec}= f_d^{3\times 3}( \Im _{out}^{enc})
    \label{Eq:dec1}
\end{equation}

\begin{equation}
    S_{d_1}^{dmr}=\Re\left ( \sum_{i=1}^{2}\beta _{n}\Big ( f^{k\times k}\left ( D_{1}^{dec} \right ) \Big ) \right ) 
    \label{Eq:mult6}
\end{equation}

\begin{equation}
    S_{d_2}^{dmr}=\Re\Bigg( \beta _{n}\Big(f^{1 \times 1}(S_{d_1}^{dmr})\Big) + S_{d_1}^{dmr} \Bigg)
    \label{Eq:mult5}
\end{equation}

\begin{equation}
    \Im _{1}^{dec}=S_{out_2}^{dmr}+\left(\beta _{n}\Big(f^{1 \times 1}(D_1^{dec})\Big)+\sum_{i=1}^{2}\beta _{n}\Big(f_d^{k\times k}(S_{d_2}^{dmr})\Big)\right)
    \label{Eq:mult4}
\end{equation}

The features $\Im _{1}^{dec} \in\mathbb {R}^{(\hat{C}\times 2)\times \frac{H}{2}\times\frac{W}{2}}$, obtained from the first decoder block, are fed into the deconvolution layer of the second decoder which, in turn, is fed to the DMRes block of the second decoder. For this purpose, the equations (Eqs.\ref{Eq:dec1}-\ref{Eq:mult4}) are repeated, and we receive $\Im_{out}^{dec}\in\mathbb{R}^{\hat{C}\times H \times W}$.
Now we evaluate the output, $X_{out}$ as given in (Eq. \ref{eqres}).
\begin{equation}
    X_{out} = \Re\Bigg(\beta_n\Big(f^{1\times 1}(\Im_{out}^{dec})\Big)\Bigg)
    \label{eqres}
\end{equation}
The feature map obtained, $X_{out}\in\mathbb{R}^{C\times H \times W}$, is processed through the dice-pixel classification layer to obtain the final binary segmentation mask, $Pred\in\mathbb{R}^{1\times H\times W}$ as in (Eq. \ref{predeq}).
\begin{equation}
    Pred = sigmoid\Big(f^{1\times1}(X_{out})\Big)
    \label{predeq}
\end{equation}

In the dual multi-residual (DMRes) blocks, we use kernels of different sizes to simultaneously capture features at varying scales on every level. This approach ensures that each feature map generated by the encoder blocks represents multi-scale features, including both high and low-frequency components. As a result, the detailed feature maps contribute to more accurate delineation of various anatomical structures. Within the DMRes blocks, we incorporate expand and squeeze blocks to accelerate convolutional operations and minimize the overall number of computations. This integration significantly enhances the model's ability to capture features at multiple scales, enabling Mini-Net to focus on both high and low-frequency features simultaneously. Additionally, the use of expand and squeeze blocks effectively reduces computational redundancy, making Mini-Net computationally efficient. 
 
\subsection{Loss Function}
We tried a bunch of popular loss functions that have shown promising performance in existing solutions for medical image segmentation tasks. Such as Dice coefficient loss given in equation \ref{eq:diceloss}, jaccard coefficient loss given in Eq.\ref{eq:jaccard}, binary cross-entropy loss given in Eq.\ref{eq:bce} and different combinations of these losses and alpha weighted loss as given in Eq.\ref{eq:alpha}. In all these equations, $Y$ represents the ground truth and $\hat{Y}$ represents the model prediction. 

The dice coefficient loss is a metric used to evaluate the overlap between the ground truth and predicted segments, particularly in image segmentation tasks. This loss function is favored for its effectiveness in addressing pixel-wise class imbalance between foreground and background regions. The dice loss can be computed as follows:
\begin{equation}
\mathcal{L}_\text{Dice} = \sum_{{I} \in \mathcal{D}}\Big(1-\frac{Y \cap \hat{Y}}{|Y|+|\hat{Y}|}\Big)^2,
\label{eq:diceloss}
\end{equation}

The Jaccard coefficient loss function, also known as the Intersection over Union (IoU) loss, has several strengths that make it a valuable choice for various machine learning tasks, particularly in image segmentation. Its strengths include robustness to class imbalance, sensitivity to object shape and boundary and direct interpretation to assess the segmentation quality. Jaccard coefficient loss can be calculated as:
\begin{equation}\label{eq:jaccard}
	\mathcal{L}_\text{Jacc.} = 1-\Big(\frac{|Y \cap \hat{Y}|}{|Y \cup \hat{Y}|}\Big)
\end{equation}

Binary cross entroy is used to measure the difference between the ground truth and the predicted binary labels. We use it in a combination with jaccard and dice loss to make the model accountable for every mislabeled pixel in the segmentation map. Binary Cross Entropy Loss:
\begin{equation}\label{eq:bce}
	\mathcal{L}_{BCE} = -{\bigg(Y\log(\hat{Y}) + (1 - Y)\log(1 - \hat{Y})\bigg)}
\end{equation}

In addition to using a combination of these popular loss functions we use a dynamic weighting mechanism for the loss functions. A dynamically weighted loss function aims to enhance the learning process by adjusting the loss function with a weight value that corresponds to the learning error of each data instance. The goal is to direct deep learning models to pay more attention to instances with larger errors, thereby improving overall performance. Alpha Weighted Loss:
\begin{equation} \label{eq:alpha}
    \mathcal{L} = \alpha \times \mathcal{L}
\end{equation}
After an extensive set of experiments on different loss functions we found out that a combination of dice coefficient loss, jaccard coefficeint and binary cross entropy loss with alpha-weighted setup gave us best segmentation results. This lead to our final loss function as:
\begin{equation} \label{eq:lossfinal}
    \mathcal{L}_{total} = \alpha(\mathcal{L}_{Dice} + \mathcal{L}_{BCE} + \mathcal{L}_{Jacc.})
\end{equation}

\section{Experiments and Implementation Details}
We conducted a comprehensive evaluation of our model, assessing its performance against the state-of-the-art using diverse datasets. The experiments involved datasets of retinal vessels, including DRIVE \cite{DRIVEdata}, STARE \cite{STAREDataset}, and CHASEDB1\cite{CHASEDataset}, as well as datasets of skin lesions such as ISIC 2016 \cite{gutman2016skin} and ISIC 2018 \cite{codella2019skin}, and the MonuSeg \cite{monusegdata} dataset. You can refer to Table \ref{tab:datasets} for specific details on these datasets, including train and test splits. All experiments were executed on a GeForce RTX 3090 GPU. For consistency between datasets, we trained Mini-Net for 100 epochs, leveraging Adam optimiser, an alpha-weighted jaccard coefficient loss function combined binary cross entropy loss given in Eq. \ref{eq:lossfinal}, and an initial learning rate set at $10^{-4}$. The utilisation of the alpha-scheduler in conjunction with the objective function proved instrumental in expediting convergence to the minima, reducing unnecessary computations, and enhancing overall training effectiveness. To enhance the efficiency of the training, we employ an early stopping approach with a patience of 4. The choice of image size and batch size varied according to each dataset's specifications, ensuring compatibility with both the dataset requirements and GPU memory limitations.

In the context of medical image segmentation, the efficacy of lighter models with fewer parameters is evident, given the inherent limitation of available datasets in the medical imaging domain. The prevalence of limited datasets makes lighter models particularly advantageous, as larger capacity models are prone to overfitting. In our approach, we start with image processing with 8 channels, gradually progressing to a maximum of 32 channels. The architectural design of our model encompasses a total of 37,685 parameters, and 36,657 are trainable. This intentional restraint in the number of parameters is a strategic choice, aligning with the need for a balanced model capacity that avoids overfitting issues commonly associated with larger models.

\begin{table*}[!htbp]
    \centering
    \caption{ Datasets used in the study.}
    \setlength\tabcolsep{8px}
    \begin{tabular}{ccccc}
        \hline
        \textbf{Application} & \textbf{Dataset} & \textbf{Image Resolution} & \textbf{Total} & \textbf{Training/Test Split} \\
        \hline
        Retinal Vessels & DRIVE \cite{DRIVEdata} & 584$\times$565 & 40 &Train: 20, Test: 20 \\
        Retinal Vessels & CHASEDB1 \cite{CHASEDataset}& 999$\times$960 & 28 &Train: 20, Test: 8 \\
        Skin Lesions & ISIC 2016 \cite{gutman2016skin}& 679$\times$453--6,748$\times$4,499 & 1,279 & Train: 900, Test: 379 \\
        Skin Lesions & ISIC 2018 \cite{codella2019skin}& 679$\times$453--6,748$\times$4,499 & 2,750 & Train: 2,000, Test: 600 \\
        Cell Nuclei & MoNuSeg \cite{monusegdata} & 1,000$\times$1,000 pixels & 44 & Train: 30, Test: 14\\ \hline
    \end{tabular}
    \label{tab:datasets}
\end{table*}

\begin{table*} [!htbp]
    \centering
    \caption{Comparison with state of the art results on the MoNuSeg \cite{monusegdata} dataset.}
        \setlength\tabcolsep{6pt}

    \begin{tabular}{lccc}
    \toprule
    \textbf{Method} & $\textbf{J}$ & $\textbf{F}_\mathbf{1}$ & \textbf{Params (M)} \\
    \midrule
    U-Net \cite{ronneberger2015u} & 0.6840 & 0.8190 & 15.56 \\
    UNet++ \cite{zhou2018unet++}  & 0.6830 & 0.8110 & 18.27 \\
    BiO-Net \cite{BiO-Net2020} & 0.7040 & 0.8240 & 15 \\
    Swin-Unet \cite{cao2023swin} & 0.6377 & 0.7769 & 82.3 \\
    UCTransNet \cite{UCTransNet2022} & 0.6668 & 0.7987 & 65.6 \\
    Proposed Mini-Net (lightweight) & \textbf{0.7056} & \textbf{0.8269} & \textbf{0.04}\\
    \bottomrule
    \end{tabular}
    
    \label{tab:MS}
\end{table*}

\begin{table*}[!htbp]
  \centering
    \setlength\tabcolsep{4pt}
  \caption{Performance comparison of Mini-Net with various SOTA methods on the skin lesion segmentation datasets ISIC 2018 \cite{codella2019skin}, and ISIC 2016 \cite{gutman2016skin}.}
    \adjustbox {max width=\textwidth}
    {
    \begin{tabular}{lcccccccccccc}
    \toprule
    \multirow{2}[5]{*}{\textbf{Method}} & \multicolumn{10}{c}{\textbf{Performance (\%)}} \\
\cmidrule{2-12}          & \multicolumn{5}{c}{\textbf{ISIC 2018}} &        & \multicolumn{5}{c}{\textbf{ISIC 2016}} \\
\cmidrule{2-6}\cmidrule{8-12}          & $Jacc$ & $F_1$  & $A_{cc}$   & $Se$   & $Sp$    &               & $Jacc$ & $F_1$  & $A_{cc}$   & $Se$   & $Sp$ \\
\cmidrule{1-6}\cmidrule{8-12}    U-Net \cite{ronneberger2015u}  & 80.09 & 86.64 & 92.52 & 85.22 & 92.09 &            & 81.38 & 88.24 & 93.31 & 87.28 & 92.88 \\
    UNet++ \cite{zhou2018unet++} & 81.62 & 87.32 & 93.72 & 88.70 & 93.96 &             & 82.81 & 89.19 & 93.88 & 88.78 & 93.52 \\
    BCDU-Net \cite{azad2019bi} & 81.10 & 85.10 & 93.70 & 78.50 & 98.20 &             & 83.43 & 80.95 & 91.78 & 78.11 & \textbf{96.20} \\
    CPFNet \cite{9049412} & 79.88 & 87.69 & 94.96 & 89.53 & 96.55 &             & 83.81 & 90.23 & 95.09 & 92.11 & 95.91 \\
    DAGAN \cite{LEI2020101716}  & 81.13 & 88.07 & 93.24 & 90.72 & 95.88 &             & 84.42 & 90.85 & 95.82 & 92.28 & 95.68 \\
    FAT-Net \cite{WU2022102327} & 82.02 & 89.03 & 95.78 & 91.00 & 96.99 &             & 85.30 & 91.59 & 96.04 & 92.59 & 96.02 \\
    AS-Net \cite{HU2022117112}  & 83.09 & 89.55 & 95.68 & 93.06 & 94.69 &            & -     & -     & -     & -     & - \\
    SLT-Net \cite{FENG2022105942} & 71.51 & 82.85 & -     & 78.85 & \textbf{99.35} &             & -     & -     & -     & -     & - \\
     Ms RED \cite{DAI2022102293} & 83.86 & 90.33 & 96.45 & 91.10 & -     &              & 87.03 & \textbf{92.66} & 96.42 & -     & - \\
    ARU-GD \cite{maji2022attention} & 84.55 & 89.16 & 94.23 & 91.42 & 96.81 &              & 85.12 & 90.83 & 94.38 & 89.86 & 94.65 \\
    Swin-Unet \cite{cao2023swin} & 82.79 & 88.98 & 96.83 & 90.10 & 97.16 &              & \textbf{87.60} & 88.94 & 96.00 & 92.27 & 95.79 \\
    \midrule
    \textbf{Mini-Net} & \textbf{89.82} & \textbf{94.47} & \textbf{96.89} & \textbf{94.22} & 97.78 &              & 87.17 & 92.45 & \textbf{96.60} & \textbf{92.51} & 95.34 \\
    \bottomrule
    \end{tabular}%
    }
  \label{tab:ISIC}%
\end{table*}%
\begin{table*}[!htbp]
  \centering
  \caption{Comparison of Mini-Net and other existing works on the DRIVE dataset \cite{DRIVEdata}. Best results are in bold, and dashes indicate unknown results.}
   \setlength\tabcolsep{6pt}
    \begin{tabular}{lccccc}
    \toprule
    \textbf{Method} & \textbf{Se} & \textbf{Sp} & \textbf{A} & $\textbf{F}_\mathbf{1}$ & \textbf{Params (M)} \\
    \midrule
    SegNet \cite{Badrinarayanan2017} & 0.7949 & 0.9738 & 0.9579  & 0.8182 & 28.40 \\
    Three-Stage FCN \cite{8476171} & 0.7631 & 0.9820 & 0.9538 & - & 20.40 \\
    Image BTS-DSN \cite{GUO2019105} & 0.7800  & 0.9806 & 0.9551 & 0.8208 & 7.80 \\
    VessNet \cite{Arsalan2019}  &   0.8022  & 0.9810  & 0.9655    & - & 9\\
    DRIU \cite{10.1007/978-3-319-46723-8_17} & 0.7855 & 0.9799 & 0.9552  & 0.8220 & 7.80 \\
    Patch BTS-DSN \cite{GUO2019105} & 0.7891 & 0.9804 & 0.9561  & 0.8249 & 7.8 \\
    DPN \cite{guo2020dpn} & 0.7934 & 0.9810 & 0.9571  & 0.818 & 3.40 \\
    MobileNet-V3 \cite{mobilenetv3} (Lightweight) & 0.8250 & 0.9771 & 0.9371 & 0.6575   & 2.50 \\
    ERFNet \cite{Romera_2018} (Lightweight) & - & - & 0.9598  & 0.7652   & 2.06 \\
    M2U-Net \cite{LaibacherWJ19_CVPRW} (Lightweight) & - & - & 0.9630  & 0.8091   & 0.55 \\
    Vessel-Net \cite{Wu2019} (Lightweight) & 0.8038 & 0.9802 & 0.9578  & -   & 1.70 \\
    MS-NFN \cite{Wu2018} (Lightweight) & 0.7844 & 0.9819 & 0.9567 & -  & 0.40 \\
    FCN \cite{OLIVEIRA2018229} (Lightweight) & 0.8039 & 0.9804 & 0.9576& - & 0.20 \\
    T-Net \cite{khan2022t} (Lightweight) & 0.8262 &  \textbf{0.9862} & {0.9697}  & 0.8269   & \textbf{0.03} \\
    ESDMR-Net (Lightweight)\cite{khan2023esdmr} & 0.8320 & 0.9832 & \textbf{0.9699} & 0.8287 & 0.70 \\
    Proposed Mini-Net(Lightweight) & \textbf{0.8370} & 0.9778 & 0.9598 & \textbf{0.8412} & 0.04\\
    \bottomrule
    \end{tabular}
    \label{tab:DRIVE}
\end{table*}

\subsection{Results and Discussion}

The exceptional performance of Mini-Net, despite its lightweight architecture, underscores its potential for broad applicability across different medical imaging modalities. The performance metrics detailed in Tables \ref{tab:MS}, \ref{tab:ISIC}, \ref{tab:DRIVE}, and \ref{tab:CHASE} consistently demonstrate Mini-Net's ability to achieve or exceed state-of-the-art results, reinforcing its robustness and efficiency.

In the context of the DRIVE dataset, as shown in Table \ref{tab:DRIVE}, Mini-Net not only achieved the highest sensitivity and $F_1$ score among lightweight models, but also maintained competitive accuracy, proving that it does not compromise performance despite its minimal parameter count. This balance between model size and performance is crucial in medical settings where computational resources are limited. It is worth mentioning that specificity of a model demonstrates the model\'s capability to identify background pixels while sensitivity demonstrates how well a model can identify foreground pixels which are actually the pixels we are interested in. Since there is a class imbalance in terms of pixel counts in medical images such that the number of background pixels are very much larger than the number of foreground pixels, it is very common for a model to show high specificity and low sensitivity. Hence, majority of the existing works have higher specificity and comparatively lower sensitivity. Nevertheless, Mini-Net displays a reasonable balance between the two metrics and is accurate enough in identifying the foreground pixels. This is because Mini-Net focuses on both the high frequency and low frequency features equally and the customized loss function makes the model capture foreground pixels accurately and learn the edges and borders more efficiently. 

For the ISIC 2016 and 2018 datasets, Mini-Net's performance, as shown in Table \ref{tab:ISIC}, was exemplary, particularly in handling high variability in image resolution and lesion appearance. This versatility is pivotal for models aimed at dermatological applications, where the morphology of the lesion can vary greatly, making consistent segmentation a challenging task. Just like on other datasets, the existing models show biased performance on skin-lesion datasets, too. The class imbalance in the dataset clearly impacts the model performance but Mini-Net again shows consistent strength in identifying the foreground pixels as well the background pixels efficiently.

Furthermore, the superior results on the CHASEDB1dataset, detailed in Table \ref{tab:CHASE}, highlight Mini-Net's proficiency in segmenting fine details such as retinal vessels, which are critical for accurate diagnostic and treatment procedures in ophthalmology. The model’s ability to finely delineate these tiny structures, often with better clarity than heavier models, could be particularly beneficial in enhancing the precision of retinal disease diagnoses.

These results collectively suggest that Mini-Net, with its innovative architecture, sets a new benchmark for lightweight models in medical image segmentation. Its impressive performance across diverse datasets indicates strong generalisability, making it a suitable choice for various real-time medical applications. 


\begin{figure*}[!htbp]
\includegraphics[width=\textwidth]{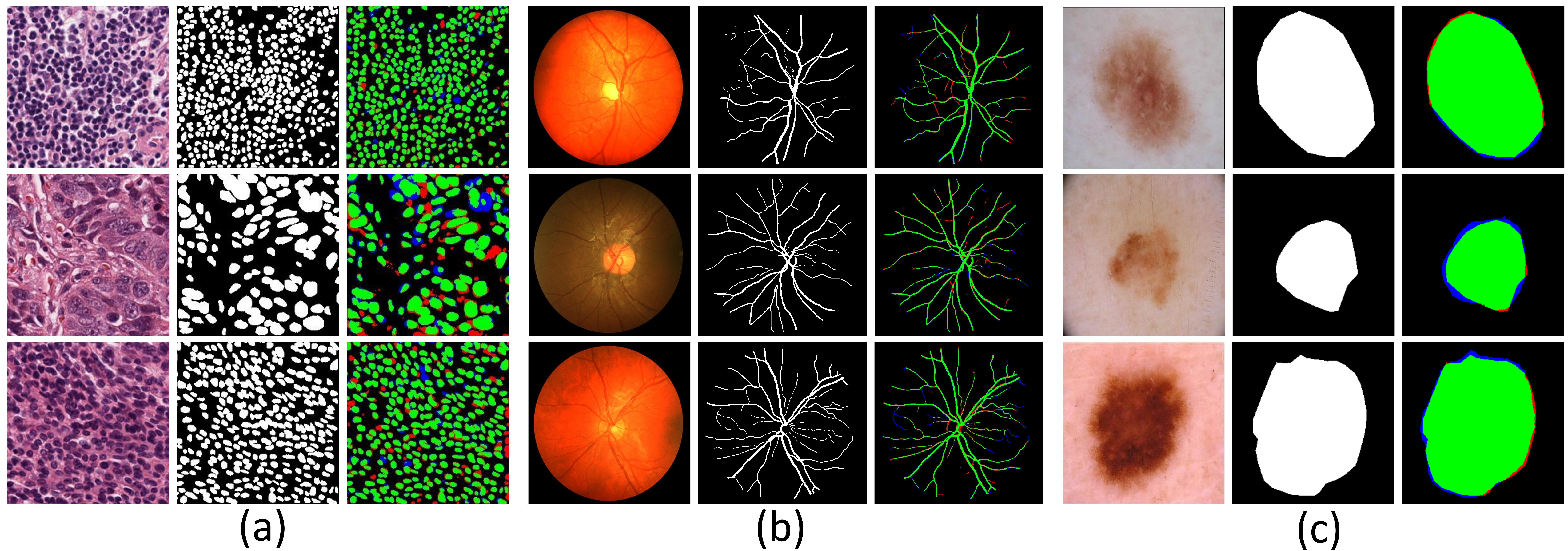}
\caption{Qualitative results of Mini-Net on sample images from (a) MonuSeg, (b) CHASE, and (c) ISIC-2018 datasets. The columns from left to right in each block represent query image, ground truth mask, and the predicted mask by Mini-Net respectively. The green and black pixels are the correctly segmented foreground and background respectively while blue pixels are the false positives and the red ones are the false negative pixels.} \label{qresult}
\end{figure*}

\begin{table}[!htbp]
  \centering
  \setlength\tabcolsep{4pt}
    \caption{Performance comparison between Mini-Net and several alternative methods on CHASEDB1 dataset \cite{CHASEDataset}.}

    \begin{tabular}{lccccc}
    \toprule
    \multirow{2}[4]{*}{\textbf{Method}} & \multicolumn{5}{c}{\textbf{Performance Measures in (\%)}} \\
\cmidrule{2-6}          & \textbf{Se} & \textbf{Sp} & \textbf{Acc} & \textbf{AUC} & \textbf{F1} \\
    \midrule
    
    SegNet \cite{Badrinarayanan2017} & 78.93 & 97.92 & 96.11 & 98.35 & 79.01 \\
    
    UNet++ \cite{zhou2018unet++} & 81.33 & 98.09 & 96.10 & 97.81 & 82.03 \\
    
    Att UNet \cite{oktay2018attention} & 80.10 & 98.04 & 96.42 & 98.40 & 80.12 \\
    
    BCD-Unet \cite{azad2019bi} & 79.41 & 98.06 & 96.07 & 97.76 & 80.22 \\

    BTS-DSN \cite{Guo2019} & 78.88 & 98.01 & 96.27 & 98.40 & 79.83 \\
    DUNet \cite{Jin2019} & 77.35 & 98.01 & 96.18 & 98.39 & 79.32 \\
    OCE-Net \cite{OCE-NET} & 81.38 & 98.24 & 96.78 & 98.72 & 81.96 \\
    Wave-Net \cite{liu2022wave} & 82.83 & 98.21 & 96.64 & -     & \textbf{83.49} \\
    
    MultiResNet \cite{maji2022attention} & 83.22 & \textbf{98.48} & 97.06 & 98.22     & 83.08 \\
    G-Net Light \cite{iqbal2022g} & 82.10 & 98.38 & 97.26 & 98.22     & 80.48 \\
    
    \midrule
    
    \textbf{Proposed Mini-Net} & \textbf{83.28} & 98.43 & \textbf{97.38} & \textbf{98.78} & 81.94 \\
    \bottomrule
    \end{tabular}%
    
  \label{tab:CHASE}%
\end{table}%
\subsection{Ablation Study}
We tried a variety of popular loss functions such as jaccard loss, binary cross-entropy loss, dice loss, a combination of these losses and an alpha-weighted version of the loss functions. As a result of extensive experiments and ablation study on loss functions, we chose the alpha-weighted sum of dice coefficient loss, binary cross-entropy and jaccard coefficient loss function. Table \ref{tab:ablation} shows the performance of our model on the ISIC-2018 dataset against different loss functions. We get the best results on the ISIC-2018 dataset with alpha-weighted sum of dice coefficient loss, binary cross entropy and Jaccard coefficient loss function which is given in Eq.\ref{eq:lossfinal}. Whereas the alpha-weighted binary cross-entropy jaccard loss function performs well with skin lesion and MonuSeg datasets, we achieved better results on retinal vessel datasets with the alpha-weighted binary cross-entropy dice loss function. It is because jaccard coefficient is more robust on object shape and boundaries than the dice coefficient loss function while these both can well handle the class imbalance between foreground and background pixels in terms of pixel count. The alpha-weighted combination of the losses work well on skin lesion and retinal vessels datasets.

\begin{table*}
\centering
\caption{Performance of model with different loss functions on ISIC-2018 dataset.}
\setlength\tabcolsep{8pt}
\begin{tabular}{llllll} \hline
Loss Function    & $Jacc$ & $F_1$  & $Acc$  & $Se$   & $Sp$   \\ \hline
Dice Loss        & 0.8787 & 0.9307 & 0.9623 & 0.9336 & 0.9608 \\
Jacc. Loss       & 0.8671 & 0.9254 & 0.9582 & 0.9183 & 0.9634 \\
BCE + Dice       & 0.8776 & 0.9294 & 0.9622 & 0.9302 & 0.9611 \\
Alpha(BCE+Dice)  & 0.8724 & 0.9266 & 0.9608 & 0.9287 & 0.9602 \\
Alpha(Jacc.)     & 0.8631 & 0.9223 & 0.9565 & 0.9218 & 0.9583 \\
Alpha(BCE+Jacc.) & 0.8814 & 0.9340 & 0.9633 & 0.9326 & 0.9631 \\
Alpha(Dice+BCE+Jacc.) & \textbf{0.8982}  & \textbf{0.9447}  & \textbf{0.9689}  & \textbf{0.9422}  & \textbf{0.9778} \\ \hline
\end{tabular}
\label{tab:ablation}
\end{table*}
\section{Conclusions}

This paper responds to the pressing need for machine learning models that can perform real-time segmentation of medical images. In addressing this need, we introduce Mini-Net, a model defined by its exceptionally lightweight framework, which is meticulously designed to support real-time segmentation tasks. Mini-Net stands out by achieving state-of-the-art results on a variety of medical image datasets, showcasing not only its effectiveness, but also its superior efficiency. With its compact design, which consists of only 37,800 parameters, Mini-Net works effectively on devices with limited memory and processing power, making it ideal for real-time medical applications.

The development of Mini-Net represents a significant advancement in the field of medical imaging, offering a solution that balances efficiency with performance. This balance is crucial for the deployment of advanced technologies in real-time settings, especially in environments where computational resources are scarce. Our comprehensive experiments across multiple datasets further highlight the model’s robust generalizability, confirming its capability to handle diverse medical imaging tasks effectively. This demonstrates Mini-Net’s potential as a transformative tool in medical diagnostics, contributing significantly to the evolution of healthcare technologies.

\bibliographystyle{plain} 
\bibliography{egbib}




\end{document}